\pdfoutput=1
\documentclass[11pt]{article}

\usepackage{graphicx}
\usepackage{cite}

\usepackage{multirow} 

\begin{document}
\begin{centering}

{\Large \textbf{Quantifying Cultural Histories via Person Networks in Wikipedia} }\\
\bigskip
Doron Goldfarb$^{1,2,3}$, Dieter Merkl$^{3}$, Maximilian Schich$^{1,2}$
\\ \medskip

{1} {\it School of Arts, Technology, and Emerging Communication\\The University of Texas at Dallas, TX, USA}\\~\\
{2} {\it Edith O'Donnell Institute of Art History\\The University of Texas at Dallas, TX, USA}\\~\\
{3} {\it Institute of Software Technology and Interactive Systems\\Vienna University of Technology, Austria}\\
\medskip
doron.goldfarb@gmail.com\\
dieter.merkl@ec.tuwien.ac.at\\
maximilian.schich@utdallas.edu

\end{centering}
\section{Introduction}
At least since Priestley's 1765 Chart of Biography \cite{priestly:chart}, large numbers of individual person records have been used to illustrate aggregate patterns of cultural history. Wikidata \cite{vrandecic:wikidata}, the structured database sister of Wikipedia, currently contains about 2.7 million explicit person records, across all language versions of the encyclopedia. These individuals, notable according to Wikipedia editing criteria, are connected via millions of hyperlinks between their respective Wikipedia articles. This situation provides us with the chance to go beyond the illustration of an idiosyncratic subset of individuals, as in the case of Priestly.
\\
\\
In this work we summarize the overlap of nationalities and occupations, based on their co-occurrence in Wikidata individuals. We construct networks of co-occurring nationalities and occupations, provide insights into their respective community structure, and apply the results to select and color chronologically structured subsets of a large network of individuals, connected by Wikipedia hyperlinks. While the imagined communities \cite{anderson:imaginedcomm} of nationality are much more discrete in terms of co-occurrence than occupations, our quantifications reveal the existing overlap of nationality as much less clear-cut than in case of occupational domains. Our work contributes to a growing body of research using biographies of notable persons to analyze cultural processes \cite{ronen:links}-\cite{goldfarb:arthistory}

\section{Method}
\begin{figure}
\begin{center}
\includegraphics[width=16cm]{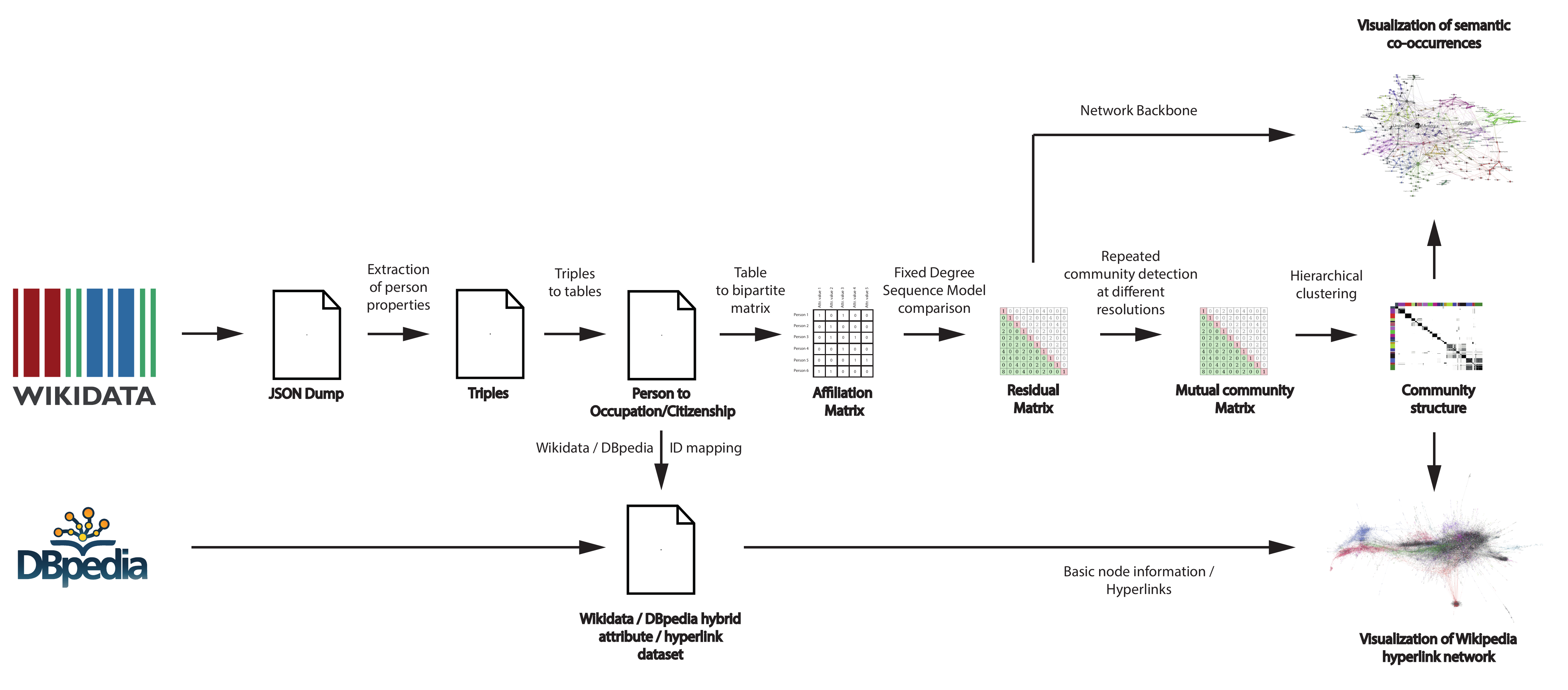}
\caption{Data processing pipeline}
\label{fig:procpip}
\end{center}

\end{figure}

\begin{figure}
\begin{center}
\includegraphics[width=10cm]{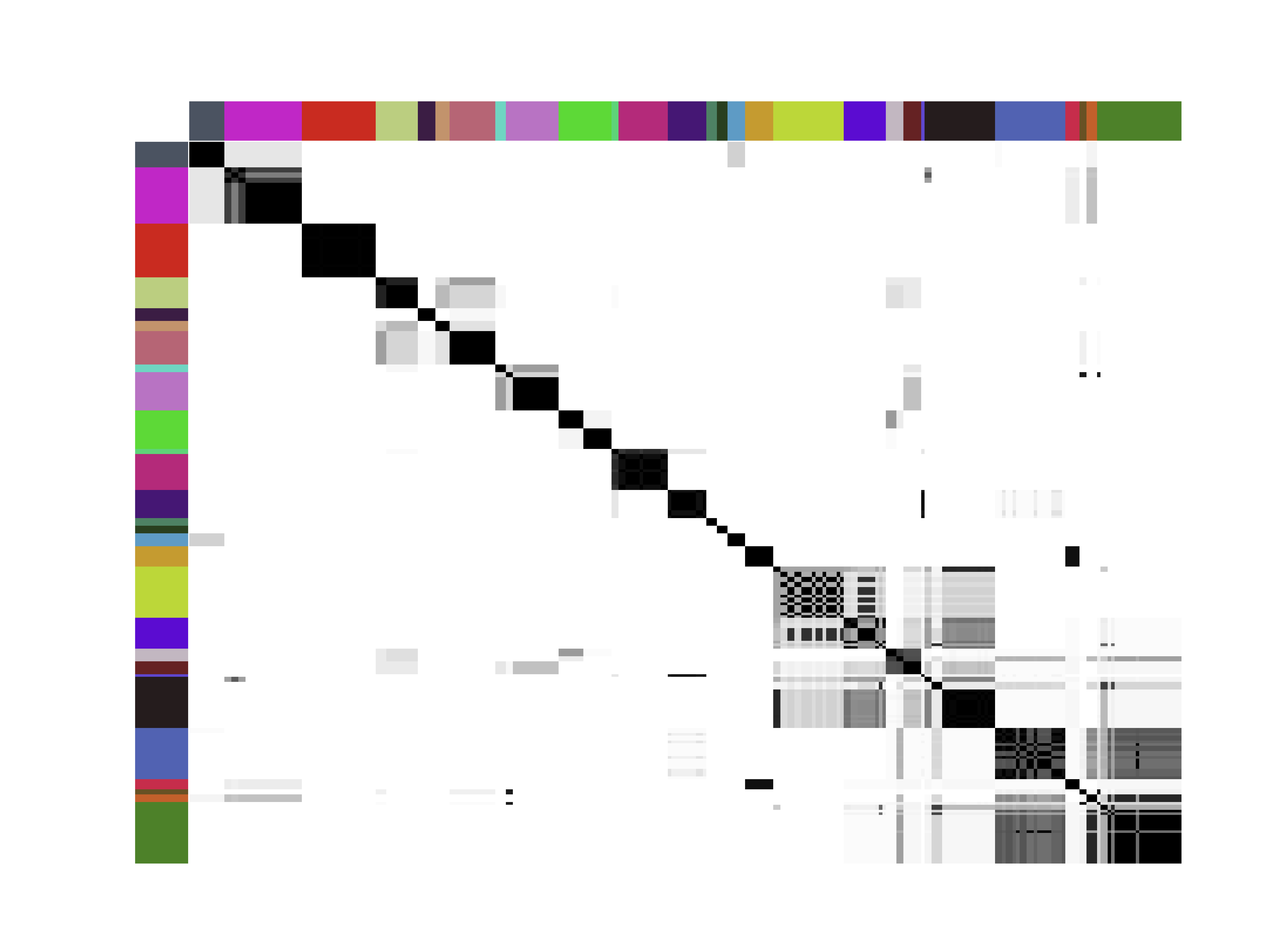}
\caption{Louvain communities of co-occurring nationalities at different resolutions}
\label{fig:citmat}
\end{center}
\end{figure}

\begin{figure}
\begin{center}
\includegraphics[width=10cm]{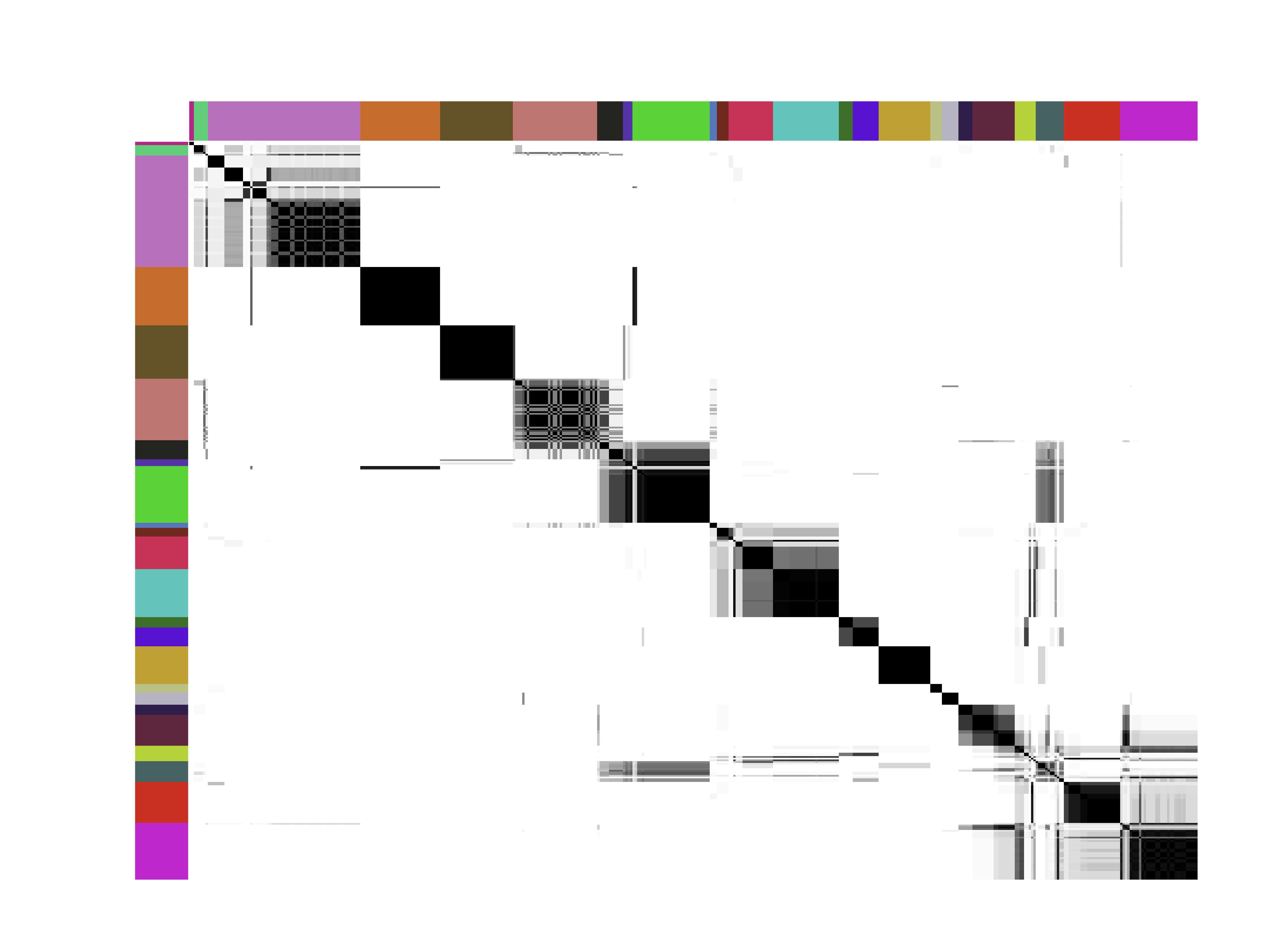}
\caption{Louvain communities of co-occurring occupations at different resolutions}
\label{fig:occmat}
\end{center}
\end{figure}

\begin{figure}
\begin{center}
\includegraphics[width=16cm]{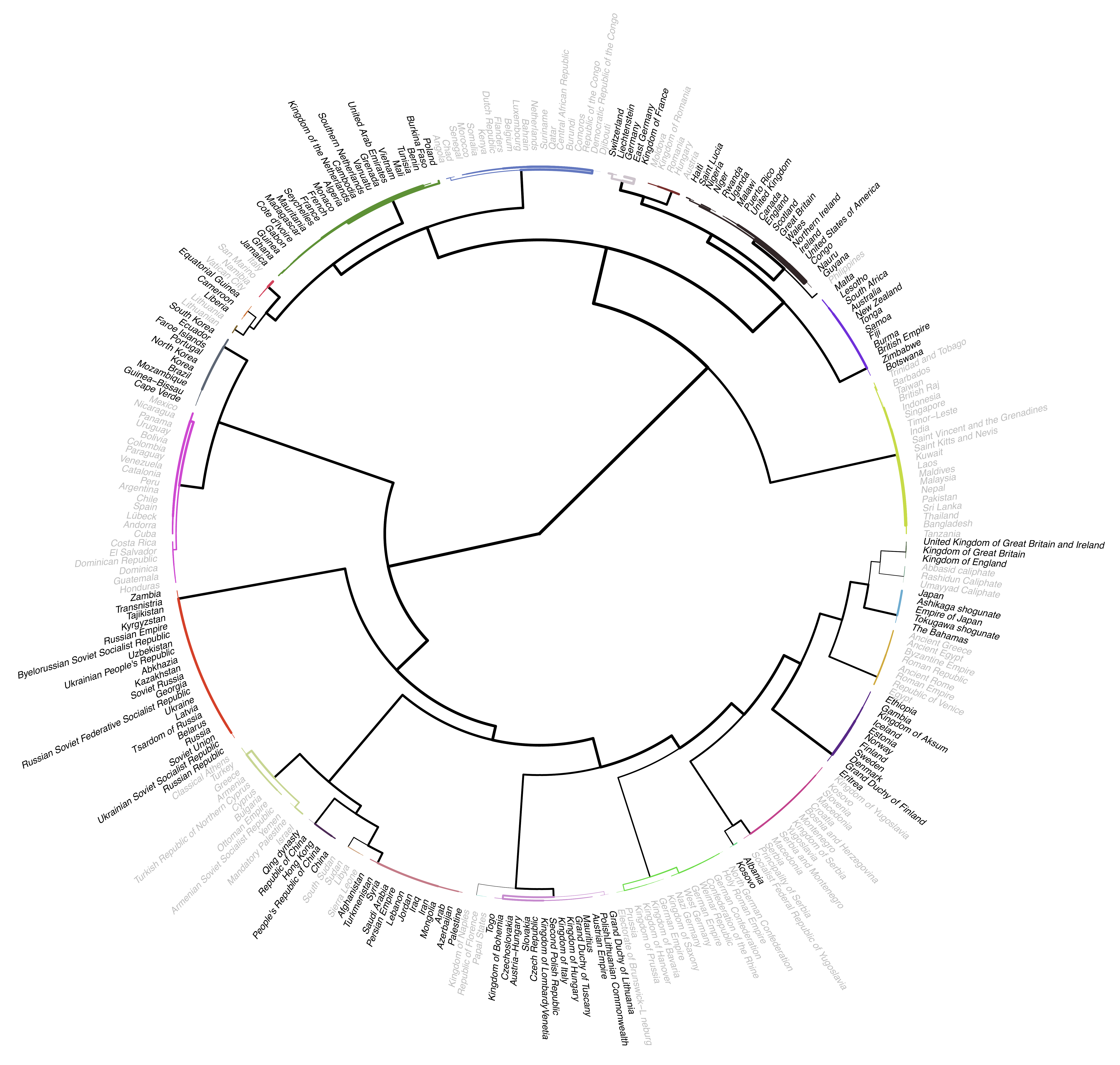}
\caption{Hierarchical clustering of communities of co-occurring nationalities}
\label{fig:citfan}
\end{center}
\end{figure}

\begin{figure}
\begin{center}
\includegraphics[width=16cm]{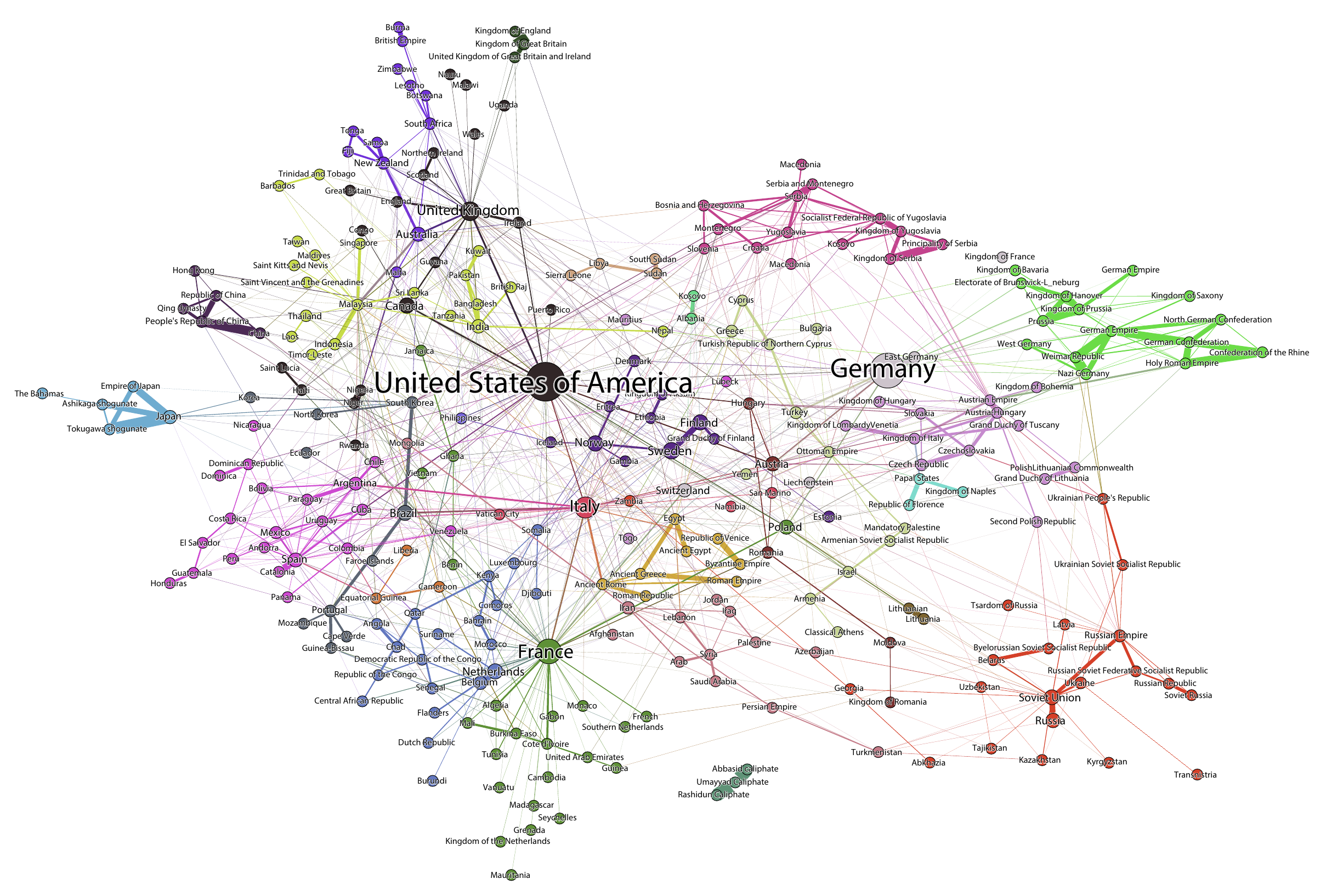}
\caption{Network of national overlap through co-occurrence, colored by community}
\label{fig:citnet}
\end{center}
\end{figure}

\begin{figure}
\begin{center}
\includegraphics[width=16cm]{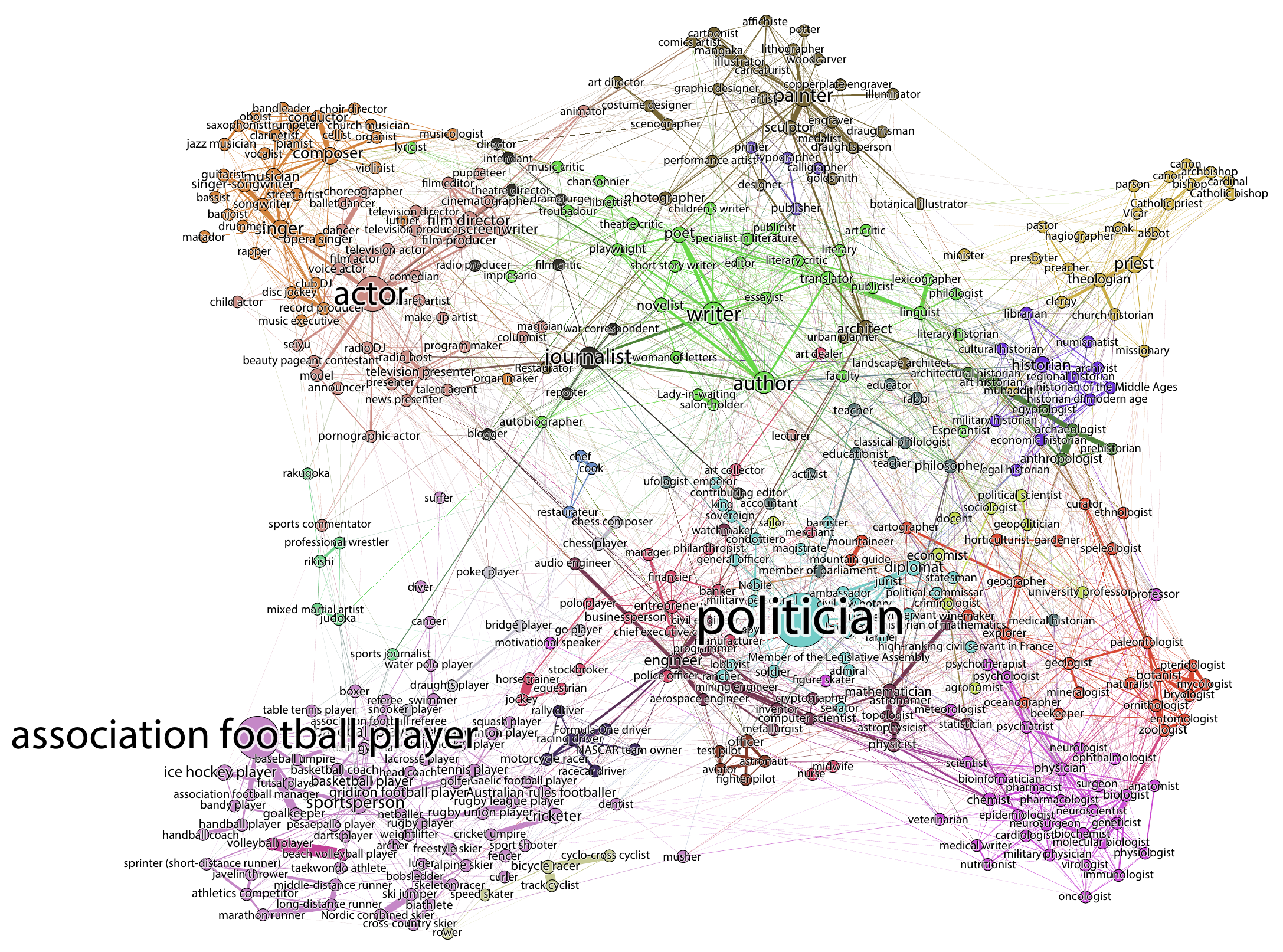}
\caption{Network of co-occurring occupations, colored by community}
\label{fig:occnet}
\end{center}
\end{figure}

\begin{figure}
\begin{center}
\includegraphics[height=12cm]{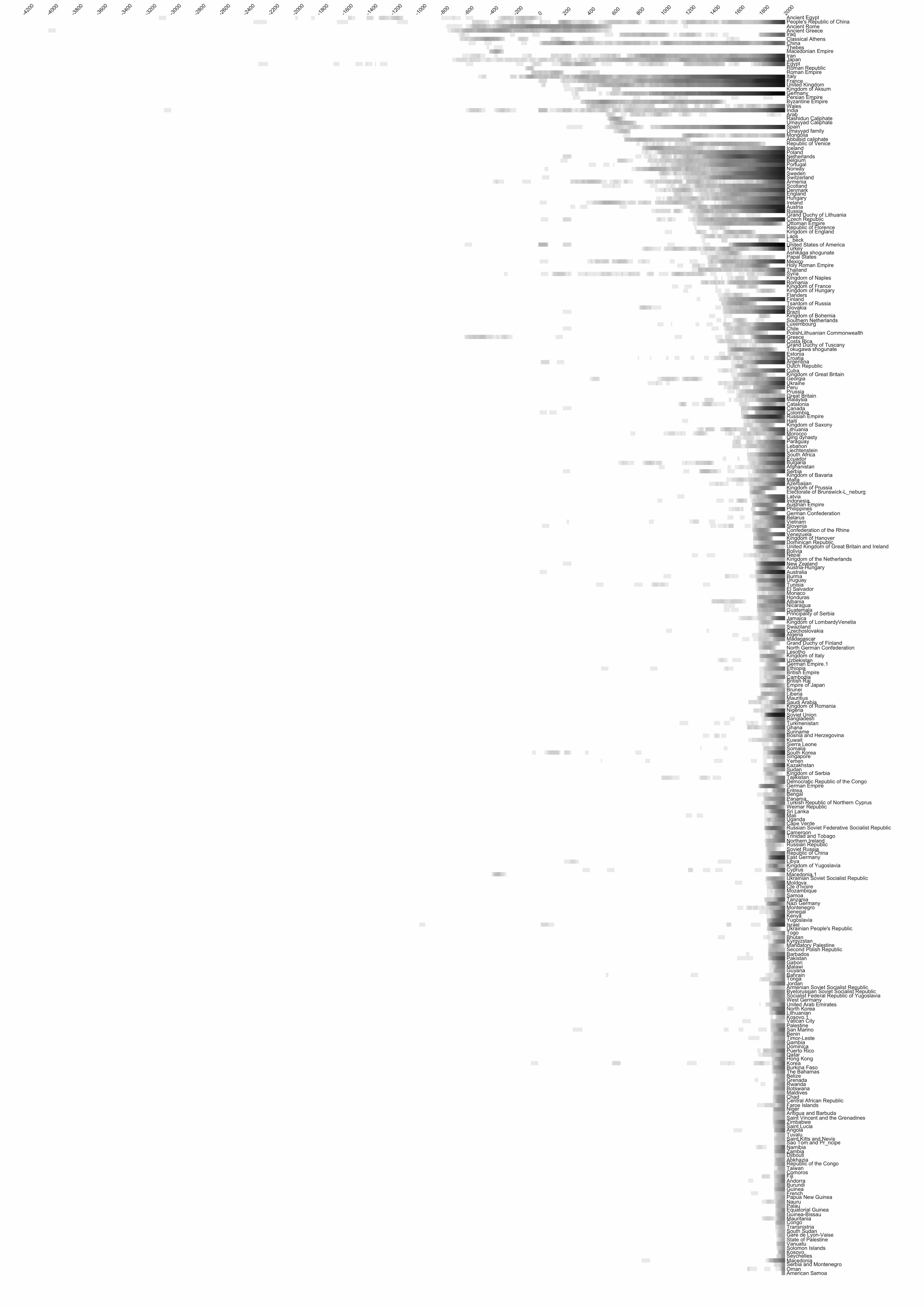}
\caption{Nationalities over time based on person life-spans}
\label{fig:cittime}
\end{center}
\end{figure}

\begin{figure}
\begin{center}
\includegraphics[height=18cm]{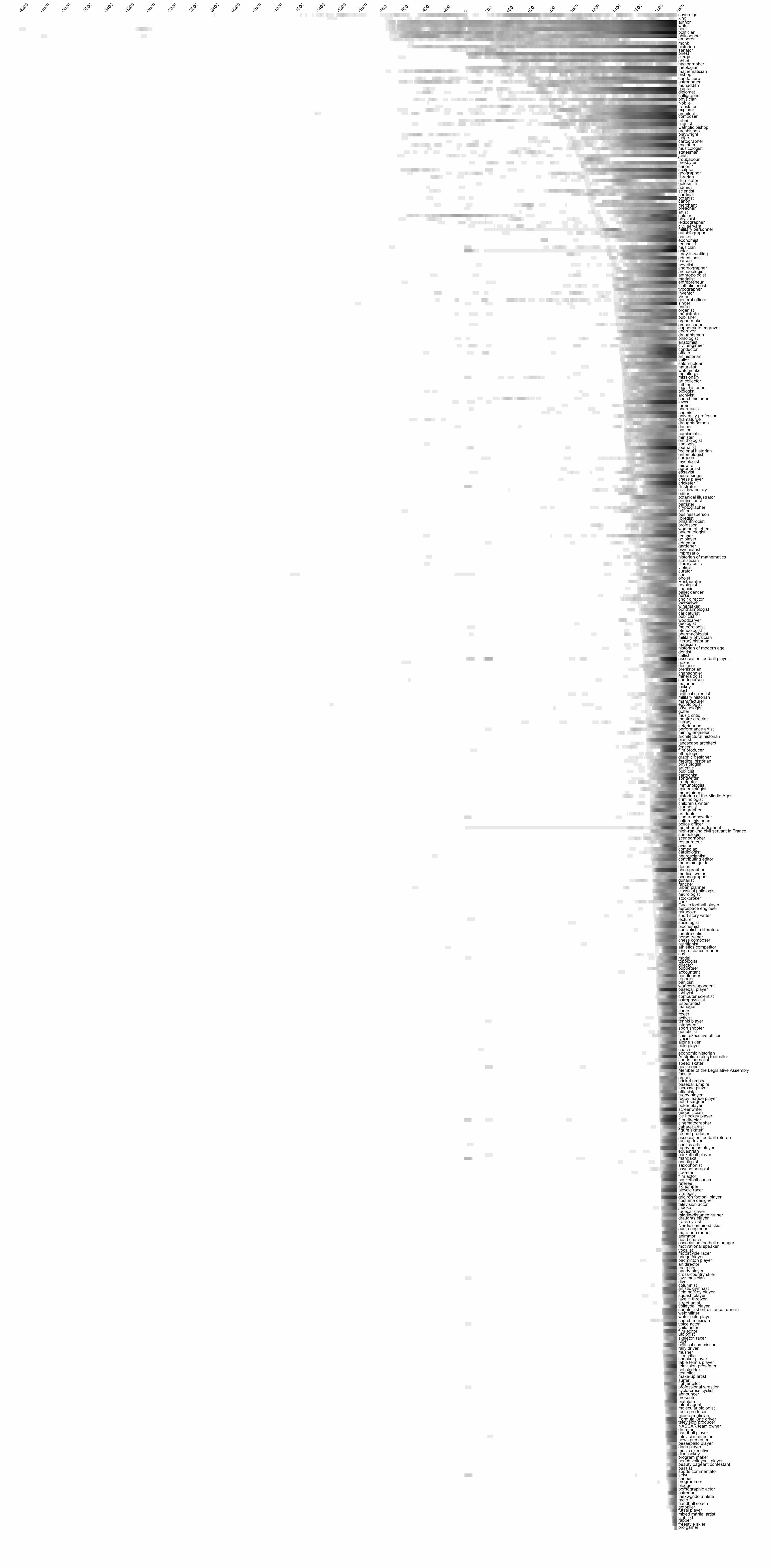}
\caption{Occupations over time based on person life-spans}
\label{fig:occtime}
\end{center}
\end{figure}

In our processing pipeline (cf. Figure \ref{fig:procpip}), we use the Wikidata Toolkit \cite{kroetsch:wikidatatoolkit} to extract 2.7 million records about humans (instances of class Q5), in the form of person -- property -- value triples, from a downloaded Wikidata json dump (09/02/2015). We focus on the properties country of citizenship (P27) and occupation (P106) (numbers see Table \ref{tab:numbers}A), restricting our analysis to nationalities with at least 10 and occupations with at least 100 occurrences. We construct and project the bipartite person-value affiliation matrices to uni-partite matrices of value-co-occurrence.
To identify relevant co-occurrences, of nationalities or occupations respectively, the projected matrices are compared against a null model. Applying an established approach \cite{zweig:bipartite}\cite{gu:fdsm}, we derive expected co-occurrence weights from an ensemble of 10,000 degree-preserving random affiliation matrices. Co-occurrences with positive Pearson residuals are considered for further analysis (numbers see Table \ref{tab:numbers}B). 
\\
\\
The resulting co-occurrence networks, with residuals as edge weights, are subsequently examined for community structure using the Louvain method \cite{blondel:louvain}\cite{traag:louvain}. Detecting communities at different granularities, we perform modularity optimization at different resolutions \cite{Reichhardt:modularity}, resulting in multiple partitions with varying numbers of communities. Using these partitions we can replace the plain co-occurrence weights in the original value-matrices with the probabilities of two values mutually co-occurring in the same community. The resulting mutual community matrix (Figures \ref{fig:citmat} and \ref{fig:occmat}) is then hierarchically clustered, with the resulting tree cut into a preset number of clusters (Figures \ref{fig:citfan} and \ref{fig:occfan}). The preset number -- 28 for nationalities, and 24 for occupations -- is based on visual inspection of repeated clusterings. 
\\
\\
Visualizations of the backbones of the co-occurrence networks (Figures \ref{fig:citnet} and \ref{fig:occnet}) show the resulting community structure in context. The network backbones are created by iteratively adding edges with the largest residuals until the maximal giant connected components (GCC) of the original networks are restored.
For comparison, we plot the occurrence of nationalities and occupations over time, ordered by their first occurrence while disregarding outliers in terms of ordering (Figures \ref{fig:cittime} and \ref{fig:occtime}). 
\\
\\
Next, the clusters of nationalities and occupations are used to partition Wikipedia biographies into national community and domain specific sub-sets. Hyperlinks connecting Wikipedia articles about individuals are obtained from DBpedia [16] and filtered to approximate contemporary relationships by excluding links between individuals with birth dates more than 75 years apart. Using hyperlinks from the English Wikipedia, we visualize the giant connected component of the partition of individuals connected to occupations in the community of "arts, architecture, crafts, and design" (Figure \ref{fig:occfish}). Colored by nationality cluster (cf. Figures \ref{fig:citmat},\ref{fig:citfan},\ref{fig:citnet}), the visualization connects 22,825 nodes with 78,447 edges. We also visualize the giant connected component of the partition of individuals connected to nationalities in the community of  "predominantly English speaking countries" (Figure \ref{fig:citfish}). Colored by occupational domain (cf. Figures \ref{fig:occmat},\ref{fig:occfan},\ref{fig:occnet}) the visualization connects 160,913 nodes with 1,004,415 edges. While the arts domain (Figure \ref{fig:occfish}) seems to reflect the established narrative of art history where a sequence of nationalities dominates at different points of time, the predominantly English speaking partition (Figure \ref{fig:citfish}) is clearly characterized by a more complex structure that excludes the construction of a simple narrative.

\section{Conclusion}
\begin{figure}
\begin{center}
\includegraphics[width=16cm]{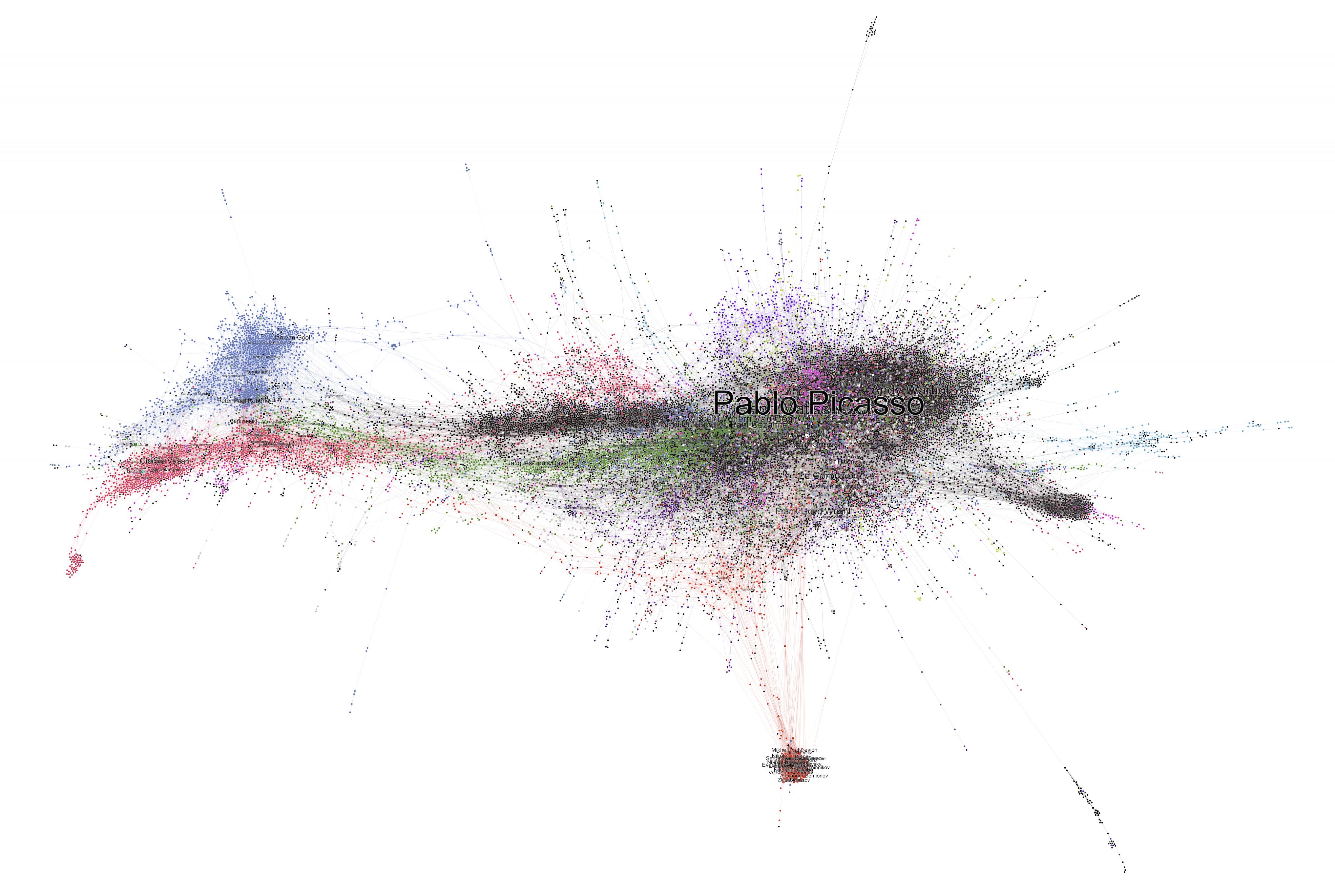}
\caption{Hyperlink network of English Wikipedia biographies having occupations in "arts, architecture, crafts and design", colored by nationality community corresponding to the colors in figures \ref{fig:citmat},\ref{fig:citfan},\ref{fig:citnet} }
\label{fig:occfish}
\end{center}
\end{figure}

In sum, we characterize networks of co-occurring nationalities and occupations related to Wikidata individuals. Our quantifications indicate that communities of nations derived from co-occurrence are much more complex than the rather clear-cut communities of occupational domain. This may be due to substantially more complex social processes leading to co-citizenship, as we observe in (post)colonial ties, due to the potentially vague concept of citizenship/nationality itself [3], as found in references to bygone and transient national constructs, or due to the considerable difference in the amount of available data (93,661 citizenship vs. 585,407 occupation co-references). Our approach can be used to group synonyms and attributions of differing granularity, occurring due to the free nature of Wikidata. 
\\
\\
Algorithmically mining occupational domains from a large set of individuals, we create an alternative to manually curated meta-domains of occupation, as used in multiple strains of recent research \cite{yu:pantheon}\cite{schich:framework}. Deriving domain specific groups of individuals directly from a crowd-sourced ecosystem, such as Wikipedia, we also provide a useful alternative (Figure \ref{fig:occfish}) to using expert curated datasets, such as the Getty Union List of Artist Names \cite{getty:vocab} as used to analyze the domain of art history in previous work \cite{goldfarb:arthistory}. Visualizing the Wikipedia hyperlink sub-networks of such domain specific groups of individuals reveals network patterns that would be obscured when using the network as a whole.

\newpage

%
%

\pagebreak

\begin{table}[htbp]
 \tiny
  \centering
  \caption{Person links to Nation/Occupation (A) and Nation/Occupation Co-occurrences (B)}
    \begin{tabular}{r|rrr|rrr}
     & \multicolumn{3}{c}{\textbf{Nationalities}} & \multicolumn{3}{c}{\textbf{Occupations}}\\
    \cline{1-7}
    \textbf{A} & \textbf{\#persons} & \textbf{\#P27 links} & \textbf{\#nationalities} & \textbf{\#persons} & \textbf{\#P106 links} & \textbf{\#occupations} \\
      \cline{1-7}
    \textbf{Raw data} & 1,318,484 & 1,366,777 & 833   & 1,363,032 & 1,706,766 & 3,419 \\
    \textbf{Reduced data} & 1,317,676 & 1,365,600 & 282   & 1,352,909 & 1,685,000 & 431 \\
    \textbf{1:1 References} & 1,271,939 & 1,271,939 & 282   & 1,099,593 & 1,099,593 & 431 \\
    \textbf{1:n References} & 45,737 & 93,661 & 282   & 253,316 & 585,407 & 430 \\
          \cline{1-7}

    \textbf{B} & \multicolumn{2}{r}{\textbf{\#co-occurrences}} & \textbf{\#nodes} & \multicolumn{2}{r}{\textbf{\#co-occurrences}} & \textbf{\#nodes} \\
              \cline{1-7}
    \textbf{All} &       & 2,100 & 282   &       & 13,846 & 430 \\
    \textbf{Positive} &       & 1,565 & 282   &       & 7,641 & 430 \\
    \textbf{Backbone} &       & 996   & 282   &       & 2,964 & 430 \\
    \end{tabular}%
  \label{tab:numbers}%
\end{table}%

\begin{figure}
\begin{center}
\includegraphics[width=14.5cm]{fig/Occupations/occupation_hierarchical.pdf}
\caption{Hierarchical clustering of communities of co-occurring occupations}
\label{fig:occfan}
\end{center}
\end{figure}

\begin{figure}
\begin{center}
\includegraphics[width=16cm]{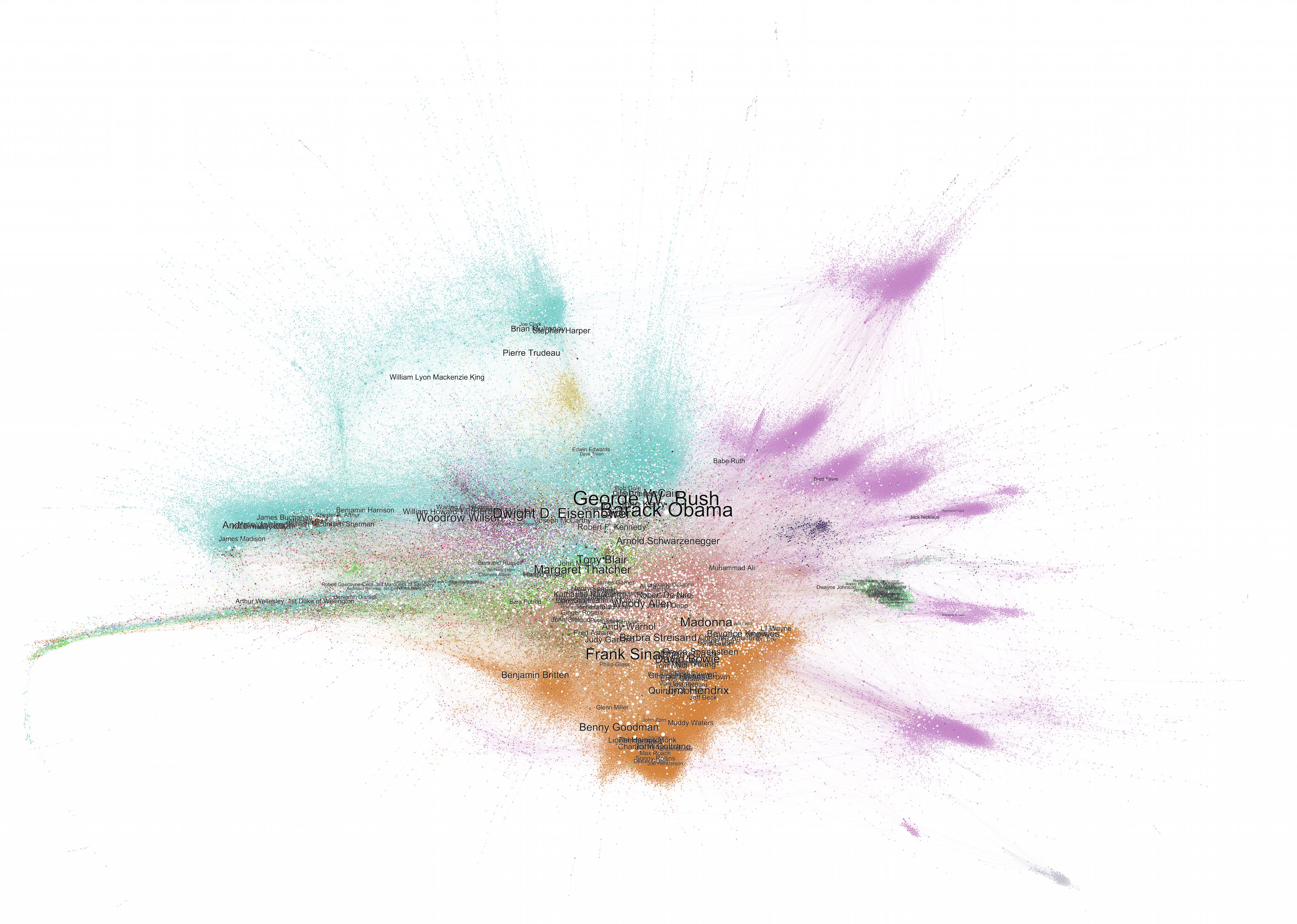}
\caption{Hyperlink network of English Wikipedia biographies with a nationality in the "predominantly english speaking" community, colored by occupation community corresponding to the colors in figures \ref{fig:occmat},\ref{fig:occfan},\ref{fig:occnet}}
\label{fig:citfish}
\end{center}
\end{figure}

\end{document}